\documentclass[a4paper]{revtex4-2}

\usepackage[utf8]{inputenc}
\usepackage[english]{babel}
\usepackage{amsmath}
\usepackage{amssymb}
\usepackage{amsfonts}
\usepackage[linktocpage=true, colorlinks=false, linkcolor = blue, citecolor = red]{hyperref}

\begin{document}
\title{Lower-dimensional Regge-Teitelboim gravity}
\author{Anton Sheykin}
\email{a.sheykin@spbu.ru}

\affiliation{St. Petersburg State University, 7-9 Universitetskaya Embankment, St. Petersburg, 199034, Russia}
\author{Agata Grechko}
\email{agatha.grechko@gmail.com}
\affiliation{Lomonosov Moscow State University, GSP-1, Leninskie Gory, Moscow, 119991, Russia}

\begin{abstract}
We study modified gravity theory known as Regge-Teitelboim approach, in which the gravity is represented by dynamics of a surface isometrically embedded in a flat bulk. We obtain some particular solutions of Regge-Teitelboim equations corresponding to a central symmetric vacuum 2+1-dimensional spacetime. In contrast with GR, this vacuum spacetime is not flat, so {it is possible for the gravitational field to exist even without matter or cosmological constant}.
\end{abstract}
\maketitle

\section{Introduction}
Although gravity was the first fundamental interaction that humankind discovered, it remains the last to be satisfactory understood. Since the appearance of general relativity, there were numerous attempts to extend or modify it. Such modifications aimed to resolve the  discrepancies of GR predictions and astrophysical ("dark matter") and cosmological ("dark energy") observations, as well as to provide a suitable framework for quantization. A satisfactory theory of gravity, as perceived by many, should meet the following criteria:
\begin{enumerate}
	\item It should adequately describe all observable processes taking place in \textit{our} universe.
	\item One should be able to find its explicit solutions corresponding to physically relevant systems and processes.
	\item It should have a sound physical and/or mathematical motivation. 
\end{enumerate}
 Unfortunately, it is quite tricky to find a theory which could meet all these criteria at once, so we end up happy when at least two of them are met. Typical example of phenomenological theories satisfying first two criteria could be MOND or mimetic gravity\cite{mimetic-review17}, while various lower-dimensional models could definitely satisfy two last ones, but fail to meet the first one because our universe is four-dimensional.
 
In this paper we consider the Regge-Teitelboim approach \cite{regge} which is a string-inspired description of gravity as a dynamics of a surface in a flat higher-dimensional bulk. It possesses non-Einsteinian solutions as it belongs to the same class as mimetic gravity (both theories appearing after a differential field transformation in GR)\cite{statja60}, but has a clear geometric interpretation. Unfortunately, the field equations of the theory (Regge-Teitelboim equations, RT) turn out to be much harder to analyze than Einstein ones\cite{deser}, and only a few particular solutions has been found (for recent developments, see \cite{statja67,statja68,Capovilla:2021nou} and references therein). For that reason in this paper we want to investigate another regime of this theory: we sacrifice the first criteria (applicability to our universe) in favor of the second one (ability to find a solution) and consider a lower-dimensional version of RT gravity. We also restrict ourselves to the central-symmetric static vacuum case.

\section{Isometric embedding of a static metric with central symmetry}
The metric of 2+1-dimensional spacetime with $SO(2)\times \mathbb{R}^1$-symmetry can be chosen as follows:
\begin{align}\label{metric}
	ds^2=A(r) dt^2-B(r)dr^2-r^2 d\phi^2.
\end{align}
Let us consider an isometric embedding of this metric into a 5-dimensional flat bulk. There are six types of $SO(2)\times \mathbb{R}^1$-symmetric surfaces which could have such symmetry \cite{statja27}. Let us restrict the consideration to four of them that can be put together in one expression\cite{statja70}: 
\begin{align}\label{y}
	\begin{split}
&y^0=k t+ \frac{h(r)}{\alpha}, \\ & y^1 = \frac{f(r)}{\alpha}  \sqrt{\varepsilon} \sin(\sqrt{\varepsilon}(\alpha t + w(r))), \\&  y^2 = \frac{f(r)}{\alpha}  \cos(\sqrt{\varepsilon}(\alpha t + w(r))),\\	
& y^3 = r \cos \phi, \\&  y^4 = r \sin \phi.
\end{split}
\end{align}
Here $k$ and $\alpha$ are constants and $\varepsilon=\pm 1$. Signature is $(\lambda, \mu \varepsilon, \mu,-1,-1)$, where $\lambda=\pm 1$ and $\mu=\pm 1$. 
The functions $f(r)$, $w(r)$ and $h(r)$ can be found using the induced metric conditions
\begin{align}\label{induced}
	\partial_\mu y^a \partial_\nu y^b \eta_{ab}=g_{\mu\nu}.
\end{align}
which give
\begin{align}\label{wf}
&f(r) = \sqrt{\frac{A-k^2\lambda}{\mu\varepsilon}}, \\ &h(r)=\int \sqrt{\frac{\alpha^2 }{\lambda A}(1-B)(A-\lambda k^2) - \frac{A'^2}{4\lambda \varepsilon A}} dr, \\	&w(r)=-\int \frac{k\lambda h'}{\mu\varepsilon f^2} dr. 
\end{align}

\section{Vacuum Regge-Teitelboim equations}
The main equations of Regge-Teitelboim approach can be obtained from EH action in which the substitution \eqref{induced} has been made. The variation w.r.t. $y^a$ gives
\begin{align}
	\partial_\mu (\sqrt{-g} G^{\mu\nu} \partial_\nu y^a) = 0.
\end{align}
Although the index $a$ can take five values, it can be shown \cite{statja18} that in 2+1-dimensional case there are only two independent RT equations.

We will need the following components of Einstein tensor corresponding to \eqref{metric}:
\begin{align}\label{G}
&	\sqrt{-g} G^{tt} = \frac{1}{2\sqrt{AB}}\frac{B'}{B}, \\& \sqrt{-g} G^{rr} = \frac{1}{2\sqrt{AB}}\frac{A'}{B}, \\& \sqrt{-g} G^{tr} = 0.
\end{align}
The equation corresponding to $a=0$ has only one term so it can be immediately integrated:
\begin{align}\label{C}
\sqrt{-g} G^{rr} \partial_r y^0 = C,  
\end{align}
where $C$ is a constant, and solved w.r.t $h(r)$:
\begin{align}\label{h}
h(r)=\int\frac{2\alpha C AB^2}{A'\sqrt{AB}}	dr
\end{align}
To obtain a compact form of another RT equation, it is convenient to take a linear combination of RT for $a=1$ and $a=2$ to get rid of the trigonometry:
\begin{align}\label{RT12}
	  \partial_t y^1 (\sqrt{g} G^{tt} \partial_t\partial_t y^2+ \partial_r(\sqrt{g} G^{rr} \partial_r y^2))-\partial_t y^2 (\sqrt{g} G^{tt} \partial_t\partial_t y^1+ \partial_r(\sqrt{g} G^{rr} \partial_r y^1))=0.
\end{align}
Plugging \eqref{wf} and \eqref{G} in \eqref{h} and \eqref{RT12}, we have the following closed system:
\begin{gather}\label{s1}
A''A'AB-A'^3 B/2 -3A'^2AB'/4+\lambda\varepsilon \alpha^2 k^2 A'B (B-1)+\varepsilon \alpha^2ABB'(\lambda k^2-A)=0\\\label{s2}
	A'^4/4-\varepsilon \alpha^2(B-1)(k^2\lambda-A)A'^2+4 \lambda \varepsilon  \alpha^2 A^2 B^3 C^2 =0.
\end{gather}

\section{Particular solutions}
The resulting system of RT equations is quite cumbersome. Let us show a few ways to obtain its particular solutions. It can be done by imposing various additional constraints.

\subsection{\textit{C=0}}
Let us suppose that the integration constant in the first RT equation \eqref{C} vanishes. It means that either $G^{rr}=0$ or $h'=0$. If $G^{rr}=0$, then $G^{tt}=0$ due to \eqref{RT12}, so $A=const$ and $B=const$ due to \eqref{G} and our spacetime is flat.   Nontrivial solutions thus must correspond to $h=const$. Solving \eqref{s2} for $B$, substituting it in \eqref{s1} and integrating, we obtain 
\begin{align}\label{A'}
	A'=\pm\frac{\alpha \sqrt{\varepsilon} A^{1/6}(A-\lambda k^2)^{5/6}}{\sqrt{A^{1/3}(A-\lambda k^2)^{2/3}-a\varepsilon (\lambda k^2-A)}}.
\end{align}  
Since there is no explicit solution of this ODE in terms of $A(r)$, we are forced to make a physical assumption. Namely, let us consider the behavior of $A$ at large $r$ and suppose that $A(r)\propto r^{n}$ with $n>0$. Then from \eqref{A'} it follows that $n-1=n/2$, so
\begin{align}
	A(r)\approx  (r/R)^2.
\end{align}
where $R$ is a constant. From \eqref{s2} it follows that 
\begin{align}
	B(r)\approx const.
\end{align}
unless one performs a fine-tuning of constants. This result raises a question whether the system \eqref{s1}-\eqref{s2} has an exact solution corresponding to such form of the metric. Let us try to find such solution.
\subsection{\textit{B(r)=const}}
In this paragraph we show that the condition $B(r)=const$ leads to quadratic dependence of $A$ on $r$. Let us assume that $B(r)=b^2=const$. Then  \eqref{s1} becomes
\begin{align}\label{w/oC}
	A''A-(A')^2/2= (1-b^2)\lambda\varepsilon \alpha^2 k^2. 
\end{align}
The general solution of \eqref{w/oC} 
\begin{align}\label{A_int}
	A(r)=a\cdot(1+r/R)^2+(1-b^2)\lambda\varepsilon \alpha^2 k^2 r^2/2,
\end{align}
{where $a$ is some constant,} would satisfy \eqref{s2} only when $b=\pm 1$ or $k=0$ {(the constant $a$ then can be absorbed in the coordinate $t$)}. The case $b= \pm 1$ is physically undesirable, since the quantity $1-b^2$ corresponds to the angular deficit in BTZ geometry \cite{Staruszkiewicz} and is related to a mass of a point source. In order to keep a nontrivial angular deficit and thus a nonzero mass of a source, one must conclude that $k=0$. Therefore, let us discuss this case in more detail.

\subsection{\textit{k=0}}
In this paragraph we will show that the condition $k=0$ alone leads to $B(r)=const$. Let us assume that $k=0$. Then the equation \eqref{s1} admits a first integral:
\begin{align}
	\frac{A'^2}{2\varepsilon \alpha^2 A B^{3/2}}+\frac{2}{\sqrt{B}}=K,
\end{align}
so we can obtain an expression for $A'$:
\begin{align}\label{d_A}
	{A'^2} = {2\varepsilon \alpha^2 A B^{3/2}} (K-2/\sqrt{B}).
\end{align}
After the substitution of \eqref{d_A} into \eqref{s2} we notice that $A$ vanishes from the equation, making it algebraic with respect to $B$. We thus conclude that $B(r)=b^2=const$, so the expressions we obtained earlier represent the only solution of \eqref{s1}-\eqref{s2} in this case. Without the loss of generality we can assume that $a=1$ (otherwise it can be absorbed in $t$) and $K=0$ in \eqref{d_A} (as both $K$ and $b$ are constants, $K$ can be absorbed in $b$). From $K=0$ it follows that $\varepsilon=-1$, otherwise the signs of \eqref{d_A} do not match. We arrive to the conclusion that the only solution of \eqref{s1}-\eqref{s2} at $k=0$ is 
\begin{align}\label{dA}
	A(r)=(1+r/R)^2, \quad B(r)=b^2.
\end{align}

\section{Shape of the surfaces}
As we saw, the RT equations are satisfied by surfaces with the following metric:
\begin{align}
	ds^2 = \left(1+\frac{r}{R}\right)^2 dt^2 - b^2 dr^2 -r^2 d\phi^2.
\end{align} Plugging the final form of the metric \eqref{dA} together with the condition $k=0$ into \eqref{wf} and solving it w.r.t. $f$ and $h$, we obtain the following form of the embedding function:
\begin{align}
	\begin{split}
	&y^0= \sqrt{-\lambda(b^2-\beta^2-1)} r, \\& y^1 = \beta (r+R) \sinh\left(\frac{{t}}{\beta R}\right), \\& y^2 =\beta {(r+R)}  \cosh\left(\frac{ {t}}{\beta R}\right),\\
	&	y^3 = r \cos \phi, \\&  y^4 = r \sin \phi. 
\end{split}
\end{align}
For $g_{tt}$ to be positively defined, one should choose $\mu=-1$, and for $g_{rr}$ to able to reach $-1$, one should choose $\lambda=1$, so the signature is $(+ + - - -)$. 

In order to obtain a more familiar form of this surface, we can redefine the coordinates and constants following Staruszkiewicz \cite{Staruszkiewicz}:
\begin{align}\label{b}
 r=\rho/b, \ \phi = b\varphi, \  \tilde{R}=R/b.
\end{align}
The constant $\beta$ is still arbitrary, but it is convenient to identify it with $b$, so the embedding function takes the form
\begin{align}\label{bb}
	\begin{split}
	&	y^0= \rho/b, \\& y^1 =  (\rho+\tilde{R}) \sinh\left(\frac{{t}}{\tilde{R}}\right), \\& y^2 =  {(\rho+\tilde{R})}  \cosh\left(\frac{ {t}}{\tilde{R}}\right),\\
	&	y^3 = \frac{\rho}{b} \cos( b\varphi), \\&  y^4 = \frac{\rho}{b} \sin (b\varphi).\end{split}
\end{align}
with metric interval
\begin{align}
	ds^2 = \left(1+\frac{\rho}{\tilde{R}}\right)^2 dt^2 - d\rho^2 -\rho^2 d\varphi^2.
\end{align}
where $\varphi\in[0,2\pi/b]$. The projection of this surface on $[y^0,y^3,y^4]$ represents a cone, so we have a singularity at $r=0$ which can be associated with the presence of a point source\cite{Willison_2011}. The angular deficit defined by $b$ is related to the mass of this source \cite{DESER1984220}.

\section{Discussion}
Above we had shown that there exist non-Einsteinian solutions of vacuum RT equations for 5-dimensional symmetric embedding of the 3-dimensional spacetime. The simplest one of them corresponds to the metric
\begin{align}
	ds^2 = \left(1+\frac{r}{R}\right)^2 dt^2 - b^2 dr^2 -r^2 d\phi^2.
\end{align}
{Such metric gives rise to a nonzero Einstein tensor, so the corresponding spacetime is non-flat and bears a gravitational field, whereas in 2+1-dimensional GR vacuum spacetimes must be flat.}
It is worth noting that in addition to the parameter $b$ (which, according to \eqref{b}, can be interpreted as a bending parameter) related to the mass of a point source there is also a constant $R$ with dimension of length. 
Its appearance is connected with the form of embedding function, whose components must have dimension of length. 
In particular, when the metric has a shift symmetry w.r.t. a cartesian coordinate (such as $t$ in the present paper), a constant parameter appears in the components of embedding function. As was shown in our previous works\cite{statja27,statja70}, there are several types of surfaces which are symmetric w.r.t. shifts of $t$. Some of them contain trigonometric or hyperbolic functions (there are the ones \eqref{y} we study here), while the others contain natural powers of $t$. All these types corresponds to representations of the translation group, whose most general form is $V(t) = \exp(\alpha W t)$, where $t$ is a translation parameter, $W$ is an arbitrary square matrix and $\alpha$ is a multiplier with dimension of inverse length. This multiplier will be present in all types of embedding function with such symmetry (except for trivial one, in which $y^0 \propto t$ and no other component depend on $t$) and could appear in the metric. In Schwarzschild case, when $[M]=L$, this parameter can be related to mass, whereas in a $2+1$-dimensional case, where $[M]=1$, it must be related to other characteristics of the manifold. It would be especially interesting to study the interplay between this parameter and the value of cosmological constant that could be added to the picture. However, the introduction of cosmological constant makes the analysis of equations much more difficult and requires additional study.

{\bf{Acknowledgments}}. The authors are grateful to S. A. Paston for useful discussions. The work is supported by RFBR Grant No.~20-01-00081.



\begin{thebibliography}{13}%
	\makeatletter
	\providecommand \@ifxundefined [1]{%
		\@ifx{#1\undefined}
	}%
	\providecommand \@ifnum [1]{%
		\ifnum #1\expandafter \@firstoftwo
		\else \expandafter \@secondoftwo
		\fi
	}%
	\providecommand \@ifx [1]{%
		\ifx #1\expandafter \@firstoftwo
		\else \expandafter \@secondoftwo
		\fi
	}%
	\providecommand \natexlab [1]{#1}%
	\providecommand \enquote  [1]{``#1''}%
	\providecommand \bibnamefont  [1]{#1}%
	\providecommand \bibfnamefont [1]{#1}%
	\providecommand \citenamefont [1]{#1}%
	\providecommand \href@noop [0]{\@secondoftwo}%
	\providecommand \href [0]{\begingroup \@sanitize@url \@href}%
	\providecommand \@href[1]{\@@startlink{#1}\@@href}%
	\providecommand \@@href[1]{\endgroup#1\@@endlink}%
	\providecommand \@sanitize@url [0]{\catcode `\\12\catcode `\$12\catcode
		`\&12\catcode `\#12\catcode `\^12\catcode `\_12\catcode `\%12\relax}%
	\providecommand \@@startlink[1]{}%
	\providecommand \@@endlink[0]{}%
	\providecommand \url  [0]{\begingroup\@sanitize@url \@url }%
	\providecommand \@url [1]{\endgroup\@href {#1}{\urlprefix }}%
	\providecommand \urlprefix  [0]{URL }%
	\providecommand \Eprint [0]{\href }%
	\providecommand \doibase [0]{https://doi.org/}%
	\providecommand \selectlanguage [0]{\@gobble}%
	\providecommand \bibinfo  [0]{\@secondoftwo}%
	\providecommand \bibfield  [0]{\@secondoftwo}%
	\providecommand \translation [1]{[#1]}%
	\providecommand \BibitemOpen [0]{}%
	\providecommand \bibitemStop [0]{}%
	\providecommand \bibitemNoStop [0]{.\EOS\space}%
	\providecommand \EOS [0]{\spacefactor3000\relax}%
	\providecommand \BibitemShut  [1]{\csname bibitem#1\endcsname}%
	\let\auto@bib@innerbib\@empty
	\bibitem [{\citenamefont {Sebastiani}\ \emph {et~al.}(2017)\citenamefont
		{Sebastiani}, \citenamefont {Vagnozzi},\ and\ \citenamefont
		{Myrzakulov}}]{mimetic-review17}%
	\BibitemOpen
	\bibfield  {author} {\bibinfo {author} {\bibfnamefont {L.}~\bibnamefont
			{Sebastiani}}, \bibinfo {author} {\bibfnamefont {S.}~\bibnamefont
			{Vagnozzi}},\ and\ \bibinfo {author} {\bibfnamefont {R.}~\bibnamefont
			{Myrzakulov}},\ }\href {https://doi.org/10.1155/2017/3156915} {\bibfield
		{journal} {\bibinfo  {journal} {Advances in High Energy Physics}\ }\textbf
		{\bibinfo {volume} {2017}},\ \bibinfo {pages} {3156915} (\bibinfo {year}
		{2017})},\ \Eprint {https://arxiv.org/abs/arXiv:1612.08661}
	{arXiv:1612.08661} \BibitemShut {NoStop}%
	\bibitem [{\citenamefont {Regge}\ and\ \citenamefont
		{Teitelboim}(1977)}]{regge}%
	\BibitemOpen
	\bibfield  {author} {\bibinfo {author} {\bibfnamefont {T.}~\bibnamefont
			{Regge}}\ and\ \bibinfo {author} {\bibfnamefont {C.}~\bibnamefont
			{Teitelboim}},\ }\bibinfo {title} {\bibfnamefont {General Relativity a la string: a progress report}}, in\ \href@noop {} {\emph {\bibinfo {booktitle} {Proceedings
				of the First Marcel Grossmann Meeting, Trieste, Italy, 1975}}},\ \bibinfo
	{editor} {edited by\ \bibinfo {editor} {\bibfnamefont {R.}~\bibnamefont
			{Ruffini}}}\ (\bibinfo {address} {North Holland, Amsterdam},\ \bibinfo {year}
	{1977})\ pp.\ \bibinfo {pages} {77--88},\ \Eprint
	{https://arxiv.org/abs/arXiv:1612.05256} {arXiv:1612.05256} \BibitemShut
	{NoStop}%
	\bibitem [{\citenamefont {Sheykin}\ \emph {et~al.}(2020)\citenamefont
		{Sheykin}, \citenamefont {Solovyev}, \citenamefont {Sukhanov},\ and\
		\citenamefont {Paston}}]{statja60}%
	\BibitemOpen
	\bibfield  {author} {\bibinfo {author} {\bibfnamefont {A.~A.}\ \bibnamefont
			{Sheykin}}, \bibinfo {author} {\bibfnamefont {D.~P.}\ \bibnamefont
			{Solovyev}}, \bibinfo {author} {\bibfnamefont {V.~V.}\ \bibnamefont
			{Sukhanov}},\ and\ \bibinfo {author} {\bibfnamefont {S.~A.}\ \bibnamefont
			{Paston}},\ }\href {https://doi.org/10.3390/sym12020240} {\bibfield
		{journal} {\bibinfo  {journal} {Symmetry}\ }\textbf {\bibinfo {volume}
			{12}},\ \bibinfo {pages} {240} (\bibinfo {year} {2020})},\ \Eprint
	{https://arxiv.org/abs/arXiv:2002.01745} {arXiv:2002.01745} \BibitemShut
	{NoStop}%
	\bibitem [{\citenamefont {Deser}\ \emph {et~al.}(1976)\citenamefont {Deser},
		\citenamefont {Pirani},\ and\ \citenamefont {Robinson}}]{deser}%
	\BibitemOpen
	\bibfield  {author} {\bibinfo {author} {\bibfnamefont {S.}~\bibnamefont
			{Deser}}, \bibinfo {author} {\bibfnamefont {F.~A.~E.}\ \bibnamefont
			{Pirani}},\ and\ \bibinfo {author} {\bibfnamefont {D.~C.}\ \bibnamefont
			{Robinson}},\ }\href {https://doi.org/10.1103/PhysRevD.14.3301} {\bibfield
		{journal} {\bibinfo  {journal} {Phys. Rev. D}\ }\textbf {\bibinfo {volume}
			{14}},\ \bibinfo {pages} {3301} (\bibinfo {year} {1976})}\BibitemShut
	{NoStop}%
	\bibitem [{\citenamefont {Paston}(2021)}]{statja67}%
	\BibitemOpen
	\bibfield  {author} {\bibinfo {author} {\bibfnamefont {S.~A.}\ \bibnamefont
			{Paston}},\ }\href {https://doi.org/10.1142/S0217732321501017} {\bibfield
		{journal} {\bibinfo  {journal} {Modern Physics Letters A}\ }\textbf {\bibinfo
			{volume} {36}},\ \bibinfo {pages} {2150101} (\bibinfo {year} {2021})},\
	\Eprint {https://arxiv.org/abs/arXiv:2006.09026} {arXiv:2006.09026}
	\BibitemShut {NoStop}%
	\bibitem [{\citenamefont {Paston}(2020)}]{statja68}%
	\BibitemOpen
	\bibfield  {author} {\bibinfo {author} {\bibfnamefont {S.~A.}\ \bibnamefont
			{Paston}},\ }\href {https://doi.org/10.3390/universe6100163} {\bibfield
		{journal} {\bibinfo  {journal} {Universe}\ }\textbf {\bibinfo {volume} {6}},\
		\bibinfo {pages} {163} (\bibinfo {year} {2020})},\ \Eprint
	{https://arxiv.org/abs/arXiv:2009.06950} {arXiv:2009.06950} \BibitemShut
	{NoStop}%
	\bibitem [{\citenamefont {Capovilla}\ \emph {et~al.}(2022)\citenamefont
		{Capovilla}, \citenamefont {Cruz},\ and\ \citenamefont
		{Rojas}}]{Capovilla:2021nou}%
	\BibitemOpen
	\bibfield  {author} {\bibinfo {author} {\bibfnamefont {R.}~\bibnamefont
			{Capovilla}}, \bibinfo {author} {\bibfnamefont {G.}~\bibnamefont {Cruz}},\
		and\ \bibinfo {author} {\bibfnamefont {E.}~\bibnamefont {Rojas}},\ }\href
	{https://doi.org/10.1142/S0218271822500080} {\bibfield  {journal} {\bibinfo
			{journal} {Int. J. Mod. Phys. D}\ }\textbf {\bibinfo {volume} {31}},\
		\bibinfo {pages} {2250008} (\bibinfo {year} {2022})},\ \Eprint
	{https://arxiv.org/abs/2110.14567} {arXiv:2110.14567 [hep-th]} \BibitemShut
	{NoStop}%
	\bibitem [{\citenamefont {Paston}\ and\ \citenamefont
		{Sheykin}(2012)}]{statja27}%
	\BibitemOpen
	\bibfield  {author} {\bibinfo {author} {\bibfnamefont {S.~A.}\ \bibnamefont
			{Paston}}\ and\ \bibinfo {author} {\bibfnamefont {A.~A.}\ \bibnamefont
			{Sheykin}},\ }\href {https://doi.org/10.1088/0264-9381/29/9/095022}
	{\bibfield  {journal} {\bibinfo  {journal} {Class. Quant. Grav.}\ }\textbf
		{\bibinfo {volume} {29}},\ \bibinfo {pages} {095022} (\bibinfo {year}
		{2012})},\ \Eprint {https://arxiv.org/abs/arXiv:1202.1204} {arXiv:1202.1204}
	\BibitemShut {NoStop}%
	\bibitem [{\citenamefont {Sheykin}\ \emph {et~al.}(2021)\citenamefont
		{Sheykin}, \citenamefont {Markov},\ and\ \citenamefont {Paston}}]{statja70}%
	\BibitemOpen
	\bibfield  {author} {\bibinfo {author} {\bibfnamefont {A.}~\bibnamefont
			{Sheykin}}, \bibinfo {author} {\bibfnamefont {M.}~\bibnamefont {Markov}},\
		and\ \bibinfo {author} {\bibfnamefont {S.}~\bibnamefont {Paston}},\ }\href
	{https://doi.org/10.1063/5.0062060} {\bibfield  {journal} {\bibinfo
			{journal} {Journal of Mathematical Physics}\ }\textbf {\bibinfo {volume}
			{62}},\ \bibinfo {pages} {102502} (\bibinfo {year} {2021})},\ \Eprint
	{https://arxiv.org/abs/arXiv:2107.00752} {arXiv:2107.00752} \BibitemShut
	{NoStop}%
	\bibitem [{\citenamefont {Paston}\ and\ \citenamefont
		{Franke}(2007)}]{statja18}%
	\BibitemOpen
	\bibfield  {author} {\bibinfo {author} {\bibfnamefont {S.~A.}\ \bibnamefont
			{Paston}}\ and\ \bibinfo {author} {\bibfnamefont {V.~A.}\ \bibnamefont
			{Franke}},\ }\href {https://doi.org/10.1007/s11232-007-0134-9} {\bibfield
		{journal} {\bibinfo  {journal} {Theor. Math. Phys.}\ }\textbf {\bibinfo
			{volume} {153}},\ \bibinfo {pages} {1582} (\bibinfo {year} {2007})},\ \Eprint
	{https://arxiv.org/abs/arXiv:0711.0576} {arXiv:0711.0576} \BibitemShut
	{NoStop}%
	\bibitem [{\citenamefont {Staruszkiewicz}(1963)}]{Staruszkiewicz}%
	\BibitemOpen
	\bibfield  {author} {\bibinfo {author} {\bibfnamefont {A.}~\bibnamefont
			{Staruszkiewicz}},\ }\href {https://www.osti.gov/biblio/4101075} {\bibfield
		{journal} {\bibinfo  {journal} {Acta Physica Polonica}\ }\textbf {\bibinfo
			{volume} {24}},\ \bibinfo {pages} {735} (\bibinfo {year} {1963})}\BibitemShut
	{NoStop}%
	\bibitem [{\citenamefont {Willison}(2011)}]{Willison_2011}%
	\BibitemOpen
	\bibfield  {author} {\bibinfo {author} {\bibfnamefont {S.}~\bibnamefont
			{Willison}},\ }\href {https://doi.org/10.1063/1.3579486} {\bibfield
		{journal} {\bibinfo  {journal} {Journal of Mathematical Physics}\ }\textbf
		{\bibinfo {volume} {52}},\ \bibinfo {pages} {042503} (\bibinfo {year}
		{2011})}\BibitemShut {NoStop}%
	\bibitem [{\citenamefont {Deser}\ \emph {et~al.}(1984)\citenamefont {Deser},
		\citenamefont {Jackiw},\ and\ \citenamefont {{'t Hooft}}}]{DESER1984220}%
	\BibitemOpen
	\bibfield  {author} {\bibinfo {author} {\bibfnamefont {S.}~\bibnamefont
			{Deser}}, \bibinfo {author} {\bibfnamefont {R.}~\bibnamefont {Jackiw}},\ and\
		\bibinfo {author} {\bibfnamefont {G.}~\bibnamefont {{'t Hooft}}},\ }\href
	{https://doi.org/https://doi.org/10.1016/0003-4916(84)90085-X} {\bibfield
		{journal} {\bibinfo  {journal} {Annals of Physics}\ }\textbf {\bibinfo
			{volume} {152}},\ \bibinfo {pages} {220} (\bibinfo {year}
		{1984})}\BibitemShut {NoStop}%
\end{thebibliography}

%

\end{document}